\documentclass [a4paper]{article}
\title{Comment on Valentini, ``De Broglie-Bohm Pilot-Wave Theory: Many Worlds in Denial?''}
\author{Harvey R Brown\thanks{Faculty of Philosophy, University of Oxford, 10 Merton Street, Oxford OX1 4JJ, U.K.; {\em harvey.brown@philosophy.ox.ac.uk.}}}  
\date{}
\begin{document}
\maketitle
\abstract{Valentini has recently defended the de Broglie-Bohm pilot-wave version of quantum theory from the claim that it is essentially Everett theory encumbered with the redundant postulation of point particles. In this note, Valentini's central arguments are questioned.

\section{Introduction}

What a privilege it is to be invited to reply to Antony Valentini's paper.\footnote{Valentini's paper \cite{valentini}, along with the present reply, will appear in \cite{saunders}.} If anyone is capable of persuading me of the plausibility of the pilot-wave picture of quantum reality, it is he. But I am not convinced that his defense of pilot-wave theory from the accusation that it is really Everett theory encumbered with otiose ontology (which Valentini calls the Claim) is successful.   In the space available, I cannot do justice to all of his arguments, so I will restrict myself to what I take to be the central ones.

On a number of occasions in his paper, Valentini stresses the philosophical point that theories should be assessed on their own terms --- that it is unfair to criticize a theory for failing to concur with assumptions that ``make sense only in rival theories''. Valentini argues that the Everett and pilot-wave pictures differ on their own terms for several reasons. 

First, the ``correct and natural viewpoint'' about pilot-wave ontology is that which Valentini attributes to de Broglie, according to which physical systems, apparatuses, people, \textit{etc}., are built from the configuration variable \textit{q}. In particular, all macroscopic, observable phenomena, including the very stuff of our mental sensations, supervene on the configurations of the punctiform corpuscles hypothesized to co-exist with the wave-function (pilot-wave). Although it is not clear that this viewpoint --- let us call it \textit{the matter assumption} --- is common to all variants of the de Broglie-Bohm approach, as Valentini admits in section 3\footnote{In Bohm's 1952 work it would seem that the role of the corpuscles in the measurement context is indexical, picking out the relevant component of the wavefunction that itself describes macroscopic physical systems; see \cite{brownwallace}.}, modern disagreements within the camp do seem to concentrate on distinct issues such as the reality or otherwise of the pilot-wave, or whether the appropriate formulation of the corpuscle dynamics is first- or second-order. Second, and more significantly, Valentini stresses the role of non-equilibrium statistics in (his own version of) pilot-wave theory. This allows him to assert that the theory is ``\textit{not} a mere alternative formulation of quantum theory''.

Finally, Valentini provides a ``counter-claim'', to the effect that the Everett picture is motivated by erroneous reasoning, and thus is unlikely to be true. In what follows, I will discuss each of these arguments in turn.

 \section{Assessing pilot-wave theory on its own terms}

A key passage in the paper occurs at the end of the discussion in section 6 of the role of decoherence in the physics of the measurement process.
\begin{quote}
Of course, \ldots, \textit{if one wishes} one may identify the flow with a set of trajectories representing parallel (approximately classical) worlds, as in the decoherence-based approach to many worlds of Saunders and Wallace. This is fair enough from a many-worlds point of view. But if we start from pilot-wave theory understood on its own terms, there is no motivation for doing so: such a step would amount to a reification of mathematical structure (assigning reality to all the trajectories asociated with the velocity field at all points in phase space). If one does so reify, one has constructed a different physical theory, with a different ontology: one may do so if one wishes, but from a pilot-wave perspective there is no special reason to take this step.
\end{quote}

The trouble is that this argument looks more like a restatement of the rival positions than a critical comparison of them, or at any rate a defense of the pilot-wave from the Claim. At the risk of belabouring this point, let me tell a story. 

Prof X has just published the latest version of his dualist philosophy of mind, which lies somewhere between solipsism and skepticism concerning other minds. Prof X hypothesizes the existence of a mental substance attached to his own person---he sees no other way of accounting for his own consciousness, and qualia in particular.  But he rejects solipsism and Berkeleian idealism, believing in the existence of an external material world including other persons. And through a questionable application of Ockham's razor, Prof X argues he can save appearances by denying mental substances, and hence (in his view) consciousness, to persons other than himself. Others may act as if they were conscious, but there was no need to go so far as to postulate that they actually are conscious.\footnote{This story may not be as contrived as it might seem. If one adopts van Fraassen's constructive empiricism, it is not entirely clear, to me at least, how any agent is supposed to avoid agnosticism about the existence of other minds.}  In her response, Prof X's arch critic, Prof Y, reiterates a point made widely in the literature, namely that to account for his own consciousness, Prof X need not appeal to dualism and the existence of mental substance; he could avail himself of a materialist, functionalist theory of mind. Were he to do this, Prof X would of course have to conclude he lives in a world with many minds. But Prof X rebuts Prof Y as follows: 
\begin{quote}
Of course, \textit{if one wishes} one may take the view that the behaviour and physical constitution of other persons are jointly evidence for the existence of other minds. This  is fair enough from the point of view of a functionalist philosophy of mind. But if we start from my dualist theory understood on its own terms, there is no motivation for doing so: such a step would amount to postulating unnecessary entities. If one does so postulate, one has constructed a different theory, with a different ontology: one may do so if one wishes, but from my dualist perspective there is no special reason to take this step.
\end{quote}

Well, functionalists like Prof Y would be forgiven for feeling a degree of frustration at this reply. Their basic claim is that dualism is unnecessary for consciousness, and therefore that Prof X's argument, rather than exploiting Ockham's razor, violates it \textit{ab initio}. If they are reasonable, functionalists should expect debate on their basic claim (which cannot be regarded as obviously true), but they will naturally regard Prof X's assertion that they are failing to assess his theory on its own terms as beside the point.

The analogy in pilot-wave theory to dualism, and in particular to mental substance, in this story is obviously the matter assumption. Why impose it? Why is it necessary within quantum mechanics to understand the nature of physical systems, apparatuses, people, \textit{etc}., in terms of configurations of hypothetical point corpuscles? If it can be shown that the wave-function or pilot-wave is structured enough to do the job, why go further? 

For many workers in quantum mechanics, the answer is clear: because without further ado unitary quantum theory faces the measurement problem. Quantum mechanics generically predicts a superposed state widely interpreted as a bizarre schizophrenia of distinct measurement outcomes. Pilot-wave theory by way of the matter assumption restores sanity, or at any rate a single definite measurement outcome.\footnote{Valentini complains in section 3 that the recent critique of pilot-wave theory by Brown and Wallace~\cite{brownwallace} is ``framed as if the measurement problem were the prime motivation for considering pilot-wave theory in the first place. As a matter of historical fact, this is false.'' Indeed, much of the historical discussion in \cite{brownwallace} purports to show that Bohm was unaware of the measurement problem in 1952. But it is hard to imagine a more significant selling-point for pilot-wave theory than its supposed ability to solve the measurement problem, whatever de Broglie's and Bohm's original motivations were.} But is it necessary to follow this route?

If the object of the exercise is to save the appearances, it is not obviously so. Everettians plausibly claim that the multiplicity of outcomes (in the sense defined by the Saunders-Wallace decoherence analysis of the superposed wavefunction) is not actually schizophrenia in any observable sense; it is consistent with experience. 

If both Everett and pilot-wave theories save the appearances, it might seem that choosing between them is a question of taste. But this would be misleading. Let us consider the case of advocates of the view that the pilot-wave is physically real (such as Valentini himself). The onus is on such advocates not just to justify the introduction of structure over and above the wavefunction on configuration space. The further onus is to explain how the matter assumption \textit{even makes sense} in the light of the possibility that the wavefunction is sufficiently highly-structured on its own to account for physical systems, apparatuses, people, \textit{etc}. After all, it is hard to see how the process of adding further degrees of freedom, hidden or otherwise, in the theory does anything to detract from the wavefunction's potency in this sense. What is really needed is an argument to the effect that the wavefunction does not have such potency as the Everettians attribute to it.\footnote{In Holland~\cite{holland}, it is merely  claimed that the wavefunction fails to have both ``form and substance'' (see the discussion in \cite{brownwallace}). A more sustained argument was offered recently by Maudlin~\cite{maudlin} in the context of any ``bare'' GRW-type theory of spontaneous collapse, which of course is just as vulnerable as Everett theory to this kind of objection---such as it is.} 

Valentini goes some way to addressing this crucial matter at two points in his paper.

At the end of section 5 of his paper, a hydrogen atom in a superposition of two `Ehrenfest' packets is discussed, and appeal is made to the possibility of a subquantum measurement designed to establish which component in the superposition is ``unoccupied''. For present purposes, the likelihood of  such a measurement being possible (see below) is largely irrelevant, the question being what the status of such a component is in itself. According to Valentini, the unoccupied component is merely ``simulating'' the approximately classical motion of the atom. Valentini further claims in section 6 that the treatment of the analogous, and more pressing, case of a superposition of non-overlapping packets representing distinct \textit{macroscopic} arrangements is conceptually just the same. But in both cases, this notion of simulation is hard to reconcile with the plausible claim that, even in pilot-wave theory taken on its own terms, the intrinsic properties of quantum systems such as mass (both inertial and gravitational), charge and magnetic moment pertain to (at least) the pilot-wave.\footnote{See \cite{brown95} and \cite{brown96}.} If in the second case the macroscopic systems involve contain human observers, and the superposition is defined relative to the appropriate decoherence basis, it is hard to see why phenomenologically the unoccupied component does not have the same status as it does in the Everett picture.\footnote{A further plausibility argument to the effect that such an unoccupied component of the superposition has the ``credentials'' to represent a \textit{bona fide} measurement outcome in standard pilot-wave theory is found in \cite{brownwallace}; the generalization of the argument to pilot-wave theory with regimes of non-equilibrium statistics is, I believe, straightforward. A critique of the credentials argument is found in \cite{lewis}.}

Further clarification is offered by Valentini in his section 7, where he question's Wallace's 2003 account \cite{wallacestructure} of the phenomenology of the wavefunction in terms of Dennett's notion of macro-objects as patterns. Valentini admits that in the quantum equilibrium regime, approximately classical experimenters 
``will encounter a phenomenological
\textit{appearance} of many worlds Ñ just as they will encounter a phenomenological
appearance of locality, uncertainty, and of quantum physics generally.''\footnote{The phrase ``phenomenological \textit{appearance} of many worlds'' is perhaps unhappy; the whole point of the Everett account of measurement is to demonstrate that the multiplicity of worlds is dynamically unavoidable but effectively unobservable, thus saving appearances! I take it, however, that what Valentini means here is that in the equilibrium regime, whatever one says about the significance of the wavefunction in Everett theory, one can say about the pilot-wave in pilot-wave theory. Valentini's position is quite different from Lewis' recent defence \cite{lewis} of pilot-wave theory, which involves simply rejecting Dennett's treatment of macro-objects as patterns (even in the equilibrium regime), but for no better reason than that it saves the day.} He again appeals to the possibility of subquantum measurements in the nonequilibrium regime to question the explanatory and predictive role of such patterns reified by Everettians. But the reality of these patterns is \textit{not} like locality and uncertainty, which are ultimately statistical notions and are supposed to depend on whether equilibrium holds. The patterns, on the other hand, are features of the wavefunction and are either there or they are not, regardless of the equilibrium condition.

\section{Non-equilibrium statistics}

A theme running throughout Valentini's paper is that pilot-wave theory cannot be a mere alternative formulation of quantum theory, or a sort of Everett theory in denial, because it allows for nonequilibrium physics. Indeed, we have just seen that Valentini effectively concedes that equilibrium pilot-wave theory is not a serious rival to Everett theory. But I argued in the last section that what is essential in the Everett picture, namely the analysis of the structural properties of the wavefunction and its ramifications for the measurement problem---what Wallace is striving to articulate in his metaphor of patterns in the context of decoherence---, is untouched by the  possibility of an additional ontology of corpuscles whether distributed in equilibrium or not. If the analysis is correct, it has implications for pilot-wave theory (or that version in which the pilot-wave is real) and bare GRW-type theories just as much as for the Everett picture. Valentini is, I think, not justified in ignoring the potency of the wavefunction in the nonequilibrium regime.

In fact, it is hard to avoid the conclusion that the very notion of nonequilibrium quantum physics is problematic. If the wavefunction is indeed potent in the relevant sense, there are strong decision-theoretic arguments to the effect that rational observers should expect Born statistics.\footnote{See the
chapter by Wallace, this volume \cite{saunders}; and strong arguments showing that rational
observers will empirically confirm the Born rule are found in the chapter by
Greaves and Myrvold, this volume \cite{saunders}. Such arguments suggest that Everett theory is more Popperian than pilot-wave theory \textit{\`{a} la} Valentini: it is more falsifiable because it rules out nonequilibrium physics in the very special regimes where Valentini thinks it is likely, but not certain.} It is not just that de Broglie-Bohm corpuscles are surplus to requirement. Their irrelevance to the issue of defining what measurement outcomes are means that their contingent distribution should pose no threat to the Born rule. Needless to say, if, as Valentini hopes, we were eventually to observe strange nonlocal phenomena associated with, say, relic particles that decoupled soon after the big bang, Everettians would have to throw in the towel. But pilot-wave theorists who treat the wavefunction as part of physical, and not just nomological, reality, should, it seems to me, be doubtful about this possibility --- unless they can show where the Everettians have gone wrong.

\section{The Counter-Claim}

Valentini claims that Everett theory, indeed Everett's own 1957 thinking, is motivated by the ``puzzle of superposition'', which in turn stems from the notion of ``eigenvalue realism''---itself allegedly based on classical reasoning.
 
The argument has several strands and is not easy to summarize concisely. But let us first consider the issue of what a measurement is in quantum mechanics. Valentini is surely right: the choice of an interaction Hamiltonian \textit{qua} measurement must ultimately be legitimized by quantum, not classical, considerations. But how does one begin? What does one mean by an observable in the first place? What makes a given interaction a ``measurement'' of that observable? And how does treatment of the measurement process tie in both with the dynamical principles in the theory and the rules governing stochastic behaviour, if any? The matter is intricate, and depends on diverse aspects of the theory. 

At the beginning of section 8, Valentini stresses that his critique of eigenvalue realism ``does not depend on pilot-wave theory'', which leaves the matter somewhat ill-defined. Exactly what version of quantum theory is in play? In the middle of section 8, Valentini states that it is ``much more likely that a new domain will be better understood in terms of a new theory based on new concepts, with its own new theory of measurement -- as shown by the example of general relativity, and indeed by the example of de Broglie's nonclassical dynamics''. Yet recall that Bohm in his 1952 version of pilot-wave theory availed himself of \textit{standard} quantum measurement theory, at least as regards the choice of interaction Hamiltonians and the definition of observables. 

Valentini, however, emphasizes in section 3 that the semantics of measurement in  pilot-wave theory (equilibrium or otherwise) is quite different from that in classical mechanics. 
But that is also largely true for orthodox quantum theory. It is widely accepted in quantum theory that generically one is not measuring what is already there, one is not revealing a pre-existing element of reality---unless, perhaps, when prior to measurement the system is in an eigenstate of the observable in question. In that special case, eigenvalue realism is very close to what in the philosophical literature is called the Eigenstate Eigenvalue Link, which in turn is very close in spirit to the 1935 Einstein-Podolsky-Rosen sufficient criterion for the existence of an element of physical reality. The common notion here is that if theory predicts that measurement of some observable will yield a certain value with probability one, then the measurement must be revealing a property of the system that was already there.  If this is right (which I doubt), then what one is supposed to infer in the generic case where the pre-measurement state is not an eigenstate is to some extent open to discussion. But it is not obvious how to avoid value-fuzziness in this case and hence a version of the puzzlement of superposition in the context of the measurement process.

Like Valentini, I think that eigenvalue realism \textit{is} questionable --- even in the absence of hidden variables. But this does not remove the the need to make sense of the superposition in the context of measurement, and Everettians do not appeal to any classical prejudices in doing so. Valentini's further argument is that Everett theory is unlikely to be true because its followers are first led to the puzzlement of superposition on the basis of eigenvalue realism.\footnote{To be fair, Valentini is, consciously or otherwise, turning on its head the 1986 Deutsch argument cited in (Valentini's) section 3 against the plausibility of pilot-wave theory based on \textit{its} allegedly suspect motivation.} Even if he were right about Everettians, the argument strikes me as unconvincing. First, successful theories are sometimes developed partly on the basis of misguided or questionable motivations. (Amongst the principal ideas driving the development of general relativity were a dubious version of Mach's principle, and the erroneous notion that the principle of general covariance represents a generalization of the relativity principle.) Second, and perhaps more pertinently, a successful theory may be developed in part to solve a long-standing conceptual problem, and in the process of doing so show precisely how the the original assumptions leading to the problem are ill-founded.

\section{Acknowledgments}
My thanks go to the Editors of this volume \cite{saunders} for the kind invitation to contribute in this way, and to  Christopher Timpson, Steve Weinstein and especially Simon Saunders for helpful discussions. I gratefully acknowledge the support of the Perimeter Institute for Theoretical Physics, where part of this work was undertaken. Research at Perimeter Institute is supported by the Government of Canada through Industry Canada and by the Province of Ontario through the Ministry of Research \& Innovation.

\end{document}